\documentclass[A4paper]{article}
\usepackage[utf8]{inputenc}
\usepackage[T1]{fontenc}
\usepackage{graphicx}
\graphicspath{{./figures/}}
\usepackage{grffile}
\usepackage{longtable}
\usepackage{wrapfig}
\usepackage{rotating}
\usepackage[normalem]{ulem}
\usepackage{textcomp}
\usepackage{amssymb}
\usepackage{capt-of}
\usepackage{url}
\usepackage{hyperref}
\usepackage{amsrefs}
\usepackage{amsmath}
\usepackage{tikz}
\usetikzlibrary{shapes,arrows}
\usepackage{stackrel}
\usepackage{tikz}
\usetikzlibrary[topaths]
\usepackage{subfig}
\usepackage{totcount}
\regtotcounter{section}
\usepackage{authblk}
\renewcommand{\Re}{\textrm{Re}}
\renewcommand{\i}{\textrm{i}}
\newcommand{\e}{\textrm{e}}

\newcommand{\w}{\omega}

\title{Stability Boundaries in Second-order Time-delayed Networks with Symmetry}
\author[1]{Diego Paolo Ferruzzo Correa\thanks{diego.ferruzzo@ufabc.edu.br}}
\author[2]{Jos\'e Roberto Castilho Piqueira\thanks{piqueira@lac.usp.br}}
\affil[1]{Centro de Engenharia, Modelagem e Ci\^encias Sociais Aplicadas da UFABC, Alameda da Universidade s/nº - S\~ao Bernardo do Campo - SP, Brazil}
\affil[2]{Escola Polit\'ecnica da Universidade de S\~ao Paulo, Av. Prof. Luciano Gualberto, 380, S\~ao Paulo - SP, Brazil}
\begin{document}
\maketitle
\section*{Abstract}
In this contribution we aim to study the stability boundaries of solutions at equilibria for a second-order oscillator networks with \(S_N\)-symmetry, we look for non-degenerate Hopf bifurcations as the time-delay between nodes increases. The remarkably simple stability criterion for synchronous solutions which, in the case of first-order self-oscillators, states that stability depends only on the sign of the coupling function derivative, is extended to a generic coupling function for second-order oscillators. As an application example, the stability boundaries for a N-node Phase-Locked Loop network is analysed.\\

\textbf{Keywords:}
oscillators network, time-delayed networks, stability, Hopf bifurcations, synchronous and asynchronous solutions.

\section{Introduction}
\label{sec:intro}
Synchronization phenomenon has attracted the attention of researchers from many areas of science for decades, mainly because it is a fundamental phenomenon observed in almost any kind of ensembles of coupled oscillators independently of their nature, from the seminal work of Strogatz \cite{Strogatz2001a} and the analysis of Arenas \cite{Arenas2008} and Acebron \cite{Acebron2005}, to more recent works which consider the lag between oscillators  \cite{Ashwin2016, Guo2015}, the continued effort to understand the underlying principles common to different kind of networks and oscillators has reached our days as a solid legacy of knowledge and tools with which we can continue exploring new frontiers,  in this sense our contribution aims to explore a gap in the literature concerning higher-order self-oscillators. By the time we write most of the research related to self-oscillators ensembles consider mainly first-order inner dynamics.

Time-delayed ensembles of second-order inner-dynamics  have been studied by the authors in a previous contribution for a particular kind of coupling function \cite{FerruzzoCorrea2015}; and symmetric bifurcations of synchronous states and the stability of small-amplitude solutions near bifurcations have been also analysed  \cite{FerruzzoCorrea2017}. Here, our first aim is to extend the well-known stability criterion presented in \cite{Earl2003}, which states that stability of synchronous solutions, in a first-order oscillator network with \(S_N\)-symmetry, depend only on the sign of the coupling function derivative, i.e., synchronous solutions are stable if \(f'(\phi_i-\phi_{j\tau})<0\); we consider a more general kind of coupling function \(f(\phi_i,\phi_{j\tau})\), introducing an stability analyses in terms of its partial derivatives for a second-order oscillator network with \(S_N\)-symmetry. This paper is divided as follows: in section \ref{sec:the-model}  the network model and some necessary definitions are introduced, in section \ref{sec:second-order} the second-order network is straight away presented and analysed, and upon it, the main results are obtained, following, in section \ref{sec:case}, a PLL network case-study is analysed; finally, conclusions and further insights are presented in section \ref{sec:conclusions}.

\section{The network model}
\label{sec:the-model}
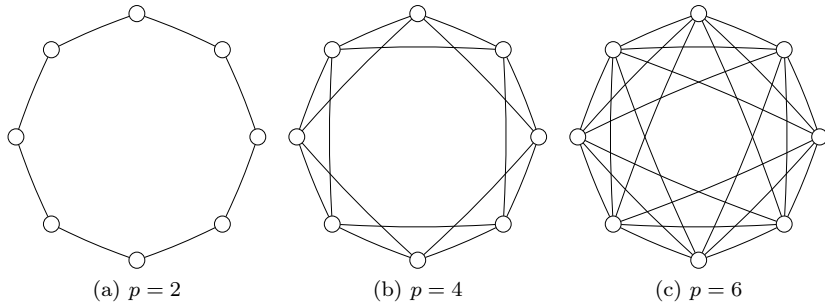
\begin{figure}[t]
\centering
\subfloat[][\(p=2\)]{
 \resizebox{3.5cm}{3.5cm}{
      \begin{tikzpicture}[transform shape]
        \foreach \x in {1,...,8}{%
          \pgfmathparse{(\x-1)*45+floor(\x/9)*22.5}
          \node[draw,circle,inner sep=0.25cm] (N-\x) at (\pgfmathresult:5.4cm){};
        };
        \path (N-1) edge[-, bend right=3] (N-2);
        \path (N-2) edge[-, bend right=3] (N-3);
        \path (N-3) edge[-, bend right=3] (N-4);
        \path (N-4) edge[-, bend right=3] (N-5);
        \path (N-5) edge[-, bend right=3] (N-6);
        \path (N-6) edge[-, bend right=3] (N-7);
        \path (N-7) edge[-, bend right=3] (N-8);
        \path (N-8) edge[-, bend right=3] (N-1);
      \end{tikzpicture}
}
}
\subfloat[][\(p=4\)]{
    \resizebox{3.5cm}{3.5cm}{
      \begin{tikzpicture}[transform shape]
        \foreach \x in {1,...,8}{%
          \pgfmathparse{(\x-1)*45+floor(\x/9)*22.5}
          \node[draw,circle,inner sep=0.25cm] (N-\x) at (\pgfmathresult:5.4cm){};
        };
        \path (N-1) edge[-, bend right=3] (N-2);
        \path (N-2) edge[-, bend right=3] (N-3);
        \path (N-3) edge[-, bend right=3] (N-4);
        \path (N-4) edge[-, bend right=3] (N-5);
        \path (N-5) edge[-, bend right=3] (N-6);
        \path (N-6) edge[-, bend right=3] (N-7);
        \path (N-7) edge[-, bend right=3] (N-8);
        \path (N-8) edge[-, bend right=3] (N-1);
        \path (N-1) edge[-, bend right=3] (N-3);
        \path (N-2) edge[-, bend right=3] (N-4);
        \path (N-3) edge[-, bend right=3] (N-5);
        \path (N-4) edge[-, bend right=3] (N-6);
        \path (N-5) edge[-, bend right=3] (N-7);
        \path (N-6) edge[-, bend right=3] (N-8);
        \path (N-7) edge[-, bend right=3] (N-1);
        \path (N-8) edge[-, bend right=3] (N-2);
      \end{tikzpicture}
}
}
\subfloat[][\(p=6\)]{
  \centering
   \resizebox{3.5cm}{3.5cm}{
      \begin{tikzpicture}[transform shape]
        \foreach \x in {1,...,8}{%
          \pgfmathparse{(\x-1)*45+floor(\x/9)*22.5}
          \node[draw,circle,inner sep=0.25cm] (N-\x) at (\pgfmathresult:5.4cm){};
        };
        \path (N-1) edge[-, bend right=3] (N-2);
        \path (N-2) edge[-, bend right=3] (N-3);
        \path (N-3) edge[-, bend right=3] (N-4);
        \path (N-4) edge[-, bend right=3] (N-5);
        \path (N-5) edge[-, bend right=3] (N-6);
        \path (N-6) edge[-, bend right=3] (N-7);
        \path (N-7) edge[-, bend right=3] (N-8);
        \path (N-8) edge[-, bend right=3] (N-1);
        \path (N-1) edge[-, bend right=3] (N-3);
        \path (N-2) edge[-, bend right=3] (N-4);
        \path (N-3) edge[-, bend right=3] (N-5);
        \path (N-4) edge[-, bend right=3] (N-6);
        \path (N-5) edge[-, bend right=3] (N-7);
        \path (N-6) edge[-, bend right=3] (N-8);
        \path (N-7) edge[-, bend right=3] (N-1);
        \path (N-8) edge[-, bend right=3] (N-2);
        \path (N-1) edge[-, bend right=3] (N-4);
        \path (N-2) edge[-, bend right=3] (N-5);
        \path (N-3) edge[-, bend right=3] (N-6);
        \path (N-4) edge[-, bend right=3] (N-7);
        \path (N-5) edge[-, bend right=3] (N-8);
        \path (N-6) edge[-, bend right=3] (N-1);
        \path (N-7) edge[-, bend right=3] (N-2);
        \path (N-8) edge[-, bend right=3] (N-3);
      \end{tikzpicture}
    }
}
13\caption{Example of an oscillator network with \(N=8\) nodes.}
\label{fig:Network}
\end{figure}

We begin by introducing the general model for a network of \(N\) identical oscillators, each of one symmetrically connected to its \(p\) closest neighbors, as shown in Figure \ref{fig:Network},
\begin{align}
\textrm{P}_M(\phi_i)=\Omega+\dfrac{K}{p-1}\sum_{j\neq i}a_{ji}f(\phi_{i},\phi_{j\tau}),~~~i=1,\ldots,N,~~2\leq p\leq N,
\label{eq:general_model}  
\end{align} 
\(\phi_i:=\phi_i(t)\) is the \(i\)-th oscillator phase, \(\phi_{j\tau}:=\phi_j(t-\tau)\) is the time-delayed phase from the \(j\)-th oscillator which is coupled through the function \(f(\phi_i,\phi_{j\tau})\); as in \cite{Earl2003} \(a_{kl}\) represents the matrix entry which is set to one if the \(k\)-th oscillator is connected to the \(l\)-th one and it is set to zero otherwise, \(\Omega\in\mathbb{R}^+\) represents the natural oscillation frequency common to all oscillators, and \(K\in\mathbb{R}^+\) the connection strength; we also introduce the linear operator \(\textrm{P}_M(\cdot)\) defined \cite{FerruzzoCorrea2013a} as:
\[
\textrm{P}_M(\cdot):=\sum_{m=1}^Mb_{m-1}\dfrac{d^{m}}{dt^{m}}(\cdot),~~~b_k\in\mathbb{R}^+, 1\leq M\in\mathbb{N},
\]
\(M\) is the  order of the oscillator; in a linear analysis with no time-delay this operator determines the characteristic polynomial of the transfer function for each oscillator whose roots characterize stability, coefficients \(b_k\) are commonly chosen to attend filtering and transient response specifications, for more details and applications see \cite{Bueno2009, Piqueira2006b, Piqueira2017}; following the Routh-Hurwitz criterion \cite{Hurwitz1895, RouthEdwardJohh2013} we restrict \(b_k\in\mathbb{R}^+\) to void instability at \(\tau=0\). 

The coupling function \(f\) is assumed to be smooth enough and surjective, such that the equilibria:
\begin{align}
\phi^*:= \left\{\phi\in\mathbb{R}, K,\Omega\in U\subset\mathbb{R}^+/\phi^{*}=f^{-1}(-\Omega/K)\right\}
\label{eq:equilibria}
\end{align}
exists, note that the \(S_N\)-symmetry is generated by the transpositions \(\pi_{ij}\),  which swaps \(\phi_i\) with \(\phi_j\), thus, it is also assume that \(f\circ\pi_{ij}=\pi_{ij}\circ f\) for all \(\pi_{ij}\).
\section{The second-order oscillator network}
\label{sec:second-order}
For \(M=2\) the model in Eq. \eqref{eq:general_model} becomes:
\begin{align}
\ddot \phi_{i} + b_0\dot{\phi}_i=\Omega+\dfrac{K}{p-1}\sum_{j\neq i}a_{ji}f(\phi_{i},\phi_{j\tau}),
\label{eq:second-order-model}  
\end{align}
here without loss of generality we have set \(b_1=1\). The linearization at equilibria gives:
\begin{align}
  \label{eq:linearization}
\ddot \phi_{i} + b_0\dot{\phi}_i-Kf'_{\phi}(\phi^*)\phi_i=\dfrac{K}{p-1}f'_{\phi\tau}(\phi^*)\sum_{j\neq i}a_{ji}\phi_{j\tau},
\end{align}
where \(f'_{\phi}(\phi^{*}):=\dfrac{\partial}{\partial \phi_{i}}f(\phi_{i},\phi_{j\tau})\bigg|_{(\phi^{*},\phi^{*})}\) and  \(f'_{\phi_{\tau}}(\phi^{*}):=\dfrac{\partial}{\partial \phi_{j\tau}}f(\phi_{i},\phi_{j\tau})\bigg|_{(\phi^{*},\phi^{*})}\). Due to the \(S_N\)-symmetry the characteristic transcendental function results in:
\begin{align}
  \begin{array}{l}    \left(\lambda^2+b_0\lambda-Kf'_{\phi}(\phi^{*})-Kf'_{\phi_{\tau}}(\phi^{*})\textrm{e}^{-\lambda\tau}\right)\\\left(\lambda^2+b_0\lambda-Kf'_{\phi}(\phi^{*})+\dfrac{K}{p-1}f'_{\phi_{\tau}}(\phi^{*})\textrm{e}^{-\lambda\tau}\right)^{N-1}=0,
    \end{array}
\label{eq:transcendental-function-1}
\end{align}
the eigenvalues of the first factor characterize the fixed-point synchronous solutions of \eqref{eq:second-order-model}, whereas the eigenvalues of the second one characterize the asynchronous solutions, note that these last eigenvalues present multiplicity \(N-1\).
We can explore the boundaries of the time-delay independent stability for synchronous and asynchronous solutions, in terms of the coupling function derivatives \(f'_{\phi}\), \(f'_{\phi_{\tau}}\) and the time-delay \(\tau\) by looking for non-degenerate Hopf bifurcations as \(\tau\) increases from zero.
For the sake of notation we rewrite \eqref{eq:transcendental-function-1}:
\begin{align}
\left(\lambda^2+\lambda-q-r\textrm{e}^{-\lambda\tau}\right)\left(\lambda^2+\lambda-q+\dfrac{r}{p-1}\textrm{e}^{-\lambda\tau}\right)^{N-1}=0,
\label{eq:transcendental-function-1a}
\end{align}
where \(q:=\dfrac{K}{b_0^2}f'_{\phi}(\phi^{*})\), \(r:=\dfrac{K}{b_0^2}f'_{\phi_{\tau}}(\phi^{*})\), \(\hat\lambda:= \dfrac{\lambda}{b_0}\) and \(\hat\tau:=b_0\tau\); for simplicity we write \(\lambda\) instead of \(\hat\lambda\) and \(\tau\) instead of \(\hat\tau\).

\subsection{Stability analysis at \(\tau=0\) and as \(\tau\to\infty\)}
\label{sec:Stabilty-analysis}
As the boundary limits of stability shall be determined by the first Hopf bifurcation in the parameter space as \(\tau\) increases, we begin by setting preliminary  conditions for \(q\) and \(r\) in order to ensure stability at \(\tau=0\).

For the synchronous solutions at \(\tau=0\), the roots of \(\lambda^2+\lambda-q-r=0\) are:
\[
\lambda_{1,2}=-\dfrac{1}{2}\pm\dfrac{1}{2}\sqrt{1+4(q+r)}
\]
thus, in order to ensure \(\Re(\lambda_{1,2})<0\) we restrict:
\begin{align}
q+r<0.
\label{eq:preliminary1a}
\end{align}
Similarly, for asynchronous solutions at \(\tau=0\), the roots of \(\lambda^2+\lambda-q+\dfrac{r}{p-1}=0\) give:
\[
\lambda_{1,2}=-\dfrac{1}{2}\pm\dfrac{1}{2}\sqrt{1+4\left(q-\dfrac{r}{p-1}\right)},
\]
and restricting
\begin{align}
q-\dfrac{r}{p-1}<0,
\label{eq:preliminary1b}
\end{align}
we ensure the stability of asynchronous solutions at \(\tau=0\).

From conditions \eqref{eq:preliminary1a} and \eqref{eq:preliminary1b} it can be seen that
\begin{align}
q<0,
\label{eq:third-preliminary-cond}
\end{align}
is a necessary but not sufficient condition to ensure  stability of synchronous and asynchronous solutions as the time-delay increases from zero.

Once condition \eqref{eq:third-preliminary-cond} is set and the time-delay is ``switched-on'', it is expected some roots to become unstable for some \(\tau>0\). It can be verified if all unstable roots remain that way as the time-delay increases further by evaluating the limit \(\tau\to\infty\) in both factors in Eq. \eqref{eq:transcendental-function-1a} assuming \(\textrm{Re}(\lambda)>0\), this implies that  roots \(\lambda_{1,2}=-\dfrac{1}{2}\pm\dfrac{1}{2}\sqrt{1+4q}\) have to have positive real part, which in turn, means \(q>0\), but this is a contradiction, once  condition \eqref{eq:third-preliminary-cond} has been set.

This observation suggest that unstable equilibria at \(\tau=0\) will remain unstable as \(\tau\to\infty\), however, the same can not be said about stable equilibria, since critical eigenvalues can cross the imaginary axis back and forth, switching stability as the time-delay increases, as proven in \cite{FerruzzoCorrea2015} for PLL oscillators.

\subsection{Hopf bifurcations}
\label{subsec:Hopf-bifs}
In order to find the boundaries of stability of solutions at equilibria (which are assumed to be stable at \(\tau=0\)) as the time-delay increases, we apply the geometric stability criteria presented in \cite{Beretta2002}, which consists in looking for eigenvalues \(\lambda=\i\omega\), with \(\omega\in\mathbb{R}^+\) along with the lowest positive time-delay \(\tau\) related for each one of the two factors in Eq. \eqref{eq:transcendental-function-1a}.

For synchronous solutions we analyze the function:
  \begin{align}
    \label{eq:pfix-char-func}
    \lambda^2+\lambda-q-r\e^{-\lambda\tau}=0,
  \end{align}
eliminating the exponential term by substituting \(\lambda=\pm\textrm{i}\omega\) we found the candidate frequency to Hopf bifurcation:
\begin{align}
  \omega=\pm\left[-\frac{1}{2}\left(2q+1\right)\pm\dfrac{1}{2}\left[\left(2q+1\right)^2-4\left(q^2-r^2\right)\right]^{1/2}\right]^{1/2},
  \label{eq:freq-Pfix}
\end{align}
and identify the region in the plane \(q\)-\(r\), with \(q<0\),  where \(\omega\in\mathbb{R}^+\):
\begin{align}
  \label{eq:conds_w_pfix}
  \left\{
  \begin{array}{ll}
    |r|>q&,-1/2\leq q<0\\
    |r|>(-q-(1/4))^{1/2}&,q<-1/2.
  \end{array}
  \right.
\end{align}

By using Euler's identity in \eqref{eq:pfix-char-func} we can obtain the time-delay corresponding to the frequency \(\omega\) found in \eqref{eq:freq-Pfix},
\begin{align}
\tau(n)=\dfrac{1}{\w}\left[\textrm{tan}^{-1}\left(\dfrac{\w}{-(\w^2+q)}\right)+2n\pi\right],~~~\tau\in\mathbb{R}^{+},~n\in\mathbb{Z},
\label{eq:tau-pfix}
\end{align}
note that for a given frequency \(\omega\) there exist infinitely many  positive time-delays of which the lowest one determines the time-delay independent stability boundary for synchronous solutions.

In order to \(\lambda=\pm\i\omega\) to be, in fact, a Hopf bifurcation at \(\tau(n)\) it is necessary to satisfy the transversality condition: \(\Re\left(\frac{\partial\lambda}{\partial\tau}\big|_{\lambda=\pm\i \omega}\right)\neq0\), which gives:
\begin{align}
-\w r\left(2\w\cos(\w\tau)+\sin(\w\tau)\right)\neq0,
\label{eq:trans-condition-1}
\end{align}
comparing \eqref{eq:freq-Pfix}, \eqref{eq:tau-pfix} and \eqref{eq:add-restrictions} we see that this condition holds if:
\begin{align}
  \label{eq:add-restrictions} 
\w\ne0~\land~|r|\ne(-q-(1/4))^{1/2}~\land~r\neq0,
\end{align}
these restrictions hold within the region defined in \eqref{eq:conds_w_pfix}.

The direction \(\delta_s\) of bifurcations is obtained directly from Eq. \eqref{eq:trans-condition-1}:
\begin{align}
\delta_s:=\textrm{sign}\left(\Re\left(\dfrac{\partial\lambda}{\partial\tau}\bigg|_{\lambda=\pm\i \omega}\right)\right)=\textrm{sign}\left(-\w r\left(2\w\cos(\w\tau)+\sin(\w\tau)\right)\right).
\label{eq:delta-s}
\end{align}

For asynchronous solutions we investigate the function:
\begin{align}
  \label{eq:pu-char-func}
    \lambda^2+\lambda-q+\dfrac{r}{p-1}\e^{-\lambda\tau}=0,
\end{align}
as before, we found the candidate frequencies to Hopf bifurcations by substituting \(\lambda=\pm\textrm{i}\omega\):
\begin{align}
  \w=\pm\left[-\frac{1}{2}\left(2 q+1\right)\pm\dfrac{1}{2}\left[\left(2q+1\right)^2-4\left(q^2-\dfrac{r^2}{(p-1)^2}\right)\right]^{1/2}\right]^{1/2},
  \label{eq:freq-Pu}
\end{align}
we see that \(\w\in\mathbb{R}^+\) within the region:
\begin{align}
  \label{eq:cond_w_PU}
  \left\{
  \begin{array}{ll}
    |r|/(p-1)>q&,-1/2\leq q<0\\
    |r|/(p-1)>(-q-(1/4))^{1/2}&,q<-1/2.
  \end{array}
  \right.
\end{align}

The time-delay for these bifurcations can also be calculated using Eq. \eqref{eq:tau-pfix}. The transversality condition for this case is:
\begin{align}
\dfrac{\w r}{p-1}\left(2\w\sin(\w\tau)+\cos(\w\tau)\right)\neq0,
\label{eq:trans-condition-2}
\end{align}
and the direction \(\delta_a\) of the bifurcations:
\begin{align}  \delta_a:=\textrm{sign}\left(\Re\left(\dfrac{\partial\lambda}{\partial\tau}\bigg|_{\lambda=\pm\i \omega}\right)\right)=\textrm{sign}\left(\dfrac{\w r}{p-1}\left(2\w\sin(\w\tau)+\cos(\w\tau)\right)\right),
  \label{eq:delta-2}
\end{align}
as it can be seen, the same conditions given in \eqref{eq:add-restrictions} must be observed for this case. 
\begin{figure}[h!]
  \begin{center}
    \includegraphics{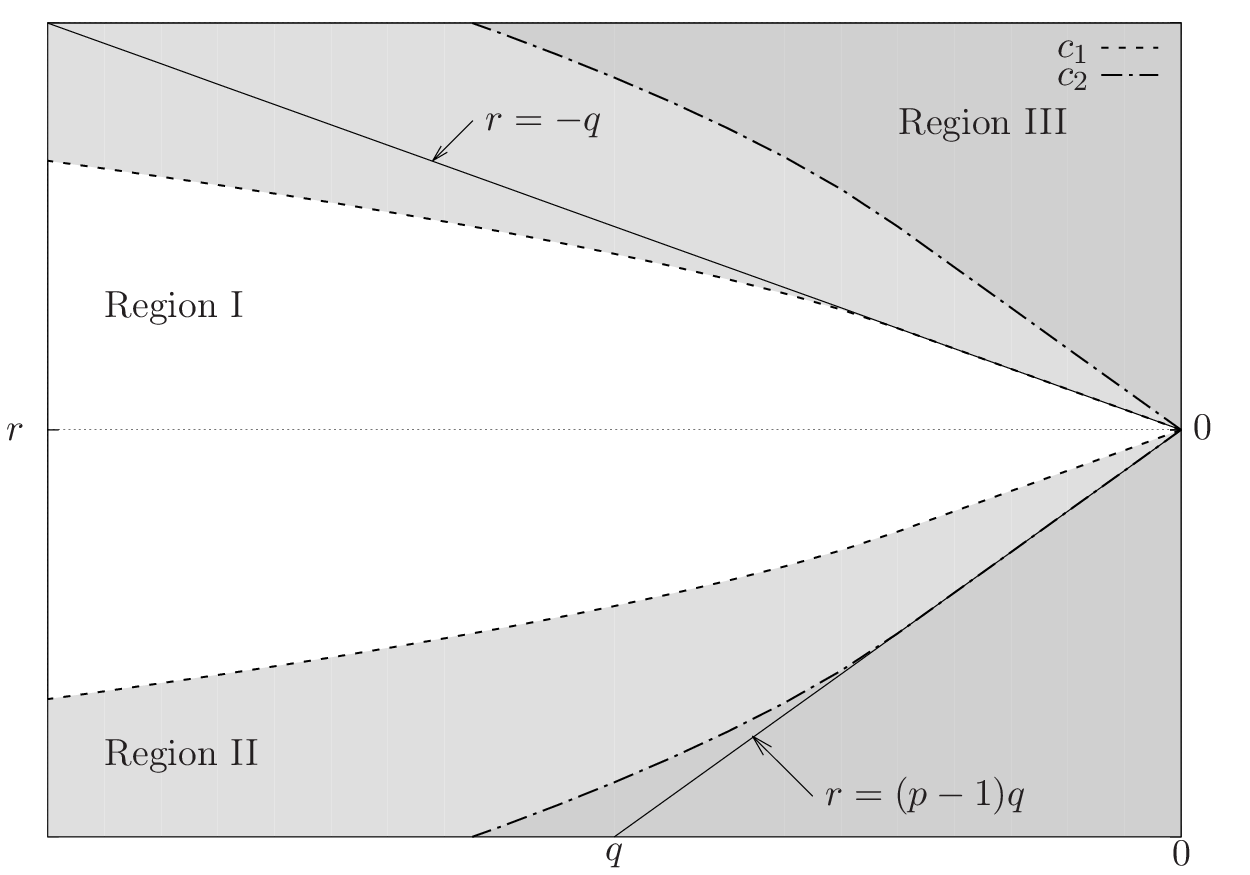}
  \end{center}
  \caption{Regions where Hopf bifurcations can emerge: solutions at equilibrium at \(\tau=0\) are stable in the region between the curves \(r=-q\) and \(r=(p-1)q\). Within Region I, bounded by curves \(c_1\), solutions are stable independently of the time-delay \(\tau\). In Region II, Hopf bifurcations for synchronous solutions emerge, and in Region III, Hopf bifurcations for both synchronous and asynchronous solutions emerge. Regions I, II and III are symmetrical with respect to the \(q\)-axis.}
    \label{fig:conditions}
\end{figure}

In Figure \ref{fig:conditions} the area delimited by the curves \(r=-q\) and \(r=-(p-1)q\) corresponds to conditions \eqref{eq:preliminary1a} and \eqref{eq:preliminary1b}, within this region roots at \(\tau=0\) are stable for both, synchronous and asynchronous solutions; the curves \(c_1\), corresponding to the boundaries of the region defined by conditions \eqref{eq:conds_w_pfix}, determine the limits where \(\omega\in\mathbb{R}^+\) for synchronous solutions, thus, in regions II and III Hopf (light and dark shaded regions) bifurcations for synchronous solutions can emerge at \(\tau(n)\) given in Eq. \eqref{eq:tau-pfix} and frequency \(\omega\) given in Eq. \eqref{eq:freq-Pfix}; similarly, curves \(c_2\), corresponding to the boundaries of conditions \eqref{eq:cond_w_PU}, delimit region III  where the frequency \(\omega\) given in Eq. \eqref{eq:freq-Pu} is real, in this region Hopf bifurcations for asynchronous solutions can emerge.

Since the region where Hopf bifurcations can emerge for asynchronous solutions is within the region where Hopf bifurcations for synchronous solutions emerge, the boundaries of this last one, curves \(c_1\) in Figure \ref{fig:conditions}, determine the upper and lower limits for \(r\neq0\), with \(q<0\), where solutions at equilibrium of system \eqref{eq:second-order-model} are time-delay independent stable. This approach to estimate the stability boundaries is conservative since, as we will see in the next section, it is possible to observe stable regions within the region where only Hopf bifurcations for synchronous solutions emerge.

\section{Case study - The PLL network}
\label{sec:case}
As an application example we analyze stability of system \eqref{eq:general_model} considering the coupling function for the PLL full-phase model:
\begin{align}
f(\phi_{i},\phi_{j\tau}):=\sin(\phi_{j\tau}-\phi_{i})+\sin(\phi_{j\tau}+\phi_{i}),
\label{eq:coupling_function}
\end{align}
the equilibria \(\phi^*=\phi^{\pm}\), defined in \eqref{eq:equilibria}, become:
\begin{align}
  \label{eq:equilibria-example}
  \begin{array}{l}
   2\phi^+=\textrm{arcsin}\left(-\Omega/K\right)\textrm{mod}~2\pi,\\
    2\phi^-=\pi-\textrm{arcsin}\left(-\Omega/K\right)\textrm{mod}~2\pi,\\
  \end{array}
\end{align}
with \(0<\Omega/K<1\), and \(q=(K/b_0^2)\left(\cos(2\phi^{\pm})-1\right)\), \(r=(K/b_0^2)\left(\cos(2\phi^{\pm})+1\right)\), with \(\cos(2\phi^{\pm})=\pm(1-\left(\Omega/K\right)^2)^{1/2}\). Stability of solutions and bifurcations for each equilibria depend on how curves:
\begin{align}
  \label{eq:curves}
  \begin{array}{ll}
    r+q=\pm K_1,&K_1:=(2K/b_0^2)(1- \left(\Omega/K\right)^2)^{1/2}\\
    r-q=K_2,&K_2:=2K/b_0^2,
    \end{array},
\end{align}
intersect regions I, II and III in Figure \ref{fig:conditions}. Note that \(K_1,K_2\in\mathbb{R}^+\) and \(K_2>K_1\).

For equilibrium \(\phi^+\) the curve \(r + q = K_1\) is outside the region where solutions are stable at \(\tau=0\), thus, this equilibrium is unstable independently of the time-delay.

\begin{figure}[h!]
    \begin{center}
      \includegraphics{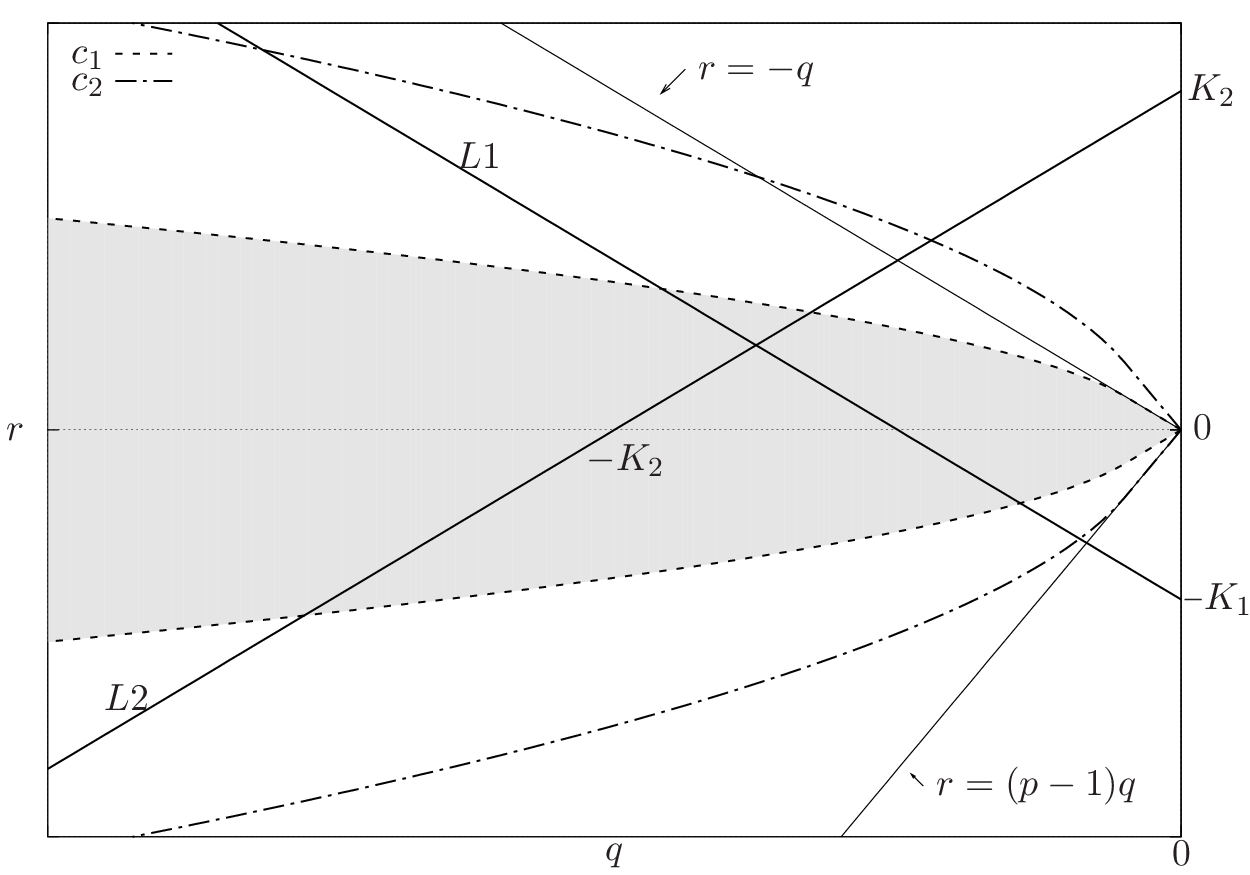}
    \end{center} 
    \caption{Intersections of curves \(L1\) and \(L2\), in Eq. \eqref{eq:curvesL1L2}, with regions I, II and III. Intersections with region I, shaded grey, determine the time-delay independent stability region.}
        \label{fig:example1}
      \end{figure}

For equilibrium \(\phi^-\), curves:
\begin{align}
    \label{eq:curvesL1L2}
    \begin{array}{l}
     L1:~r+q = -K_1\\
     L2:~r-q = K_2,
    \end{array}
\end{align}
 are shown in Figure \ref{fig:example1}  upon regions I, II and III, the time-delay independent stability region is now shaded gray, clearly, the intersections of \(L1\) and \(L2\) with the boundaries of this region, determine the set of values of parameters \(K\), \(b_0\) and \(\Omega\) for which synchronous and asynchronous solutions are stable independently of the time-delay.

\begin{figure}[ht!]
  \centering
  \includegraphics[scale=0.75]{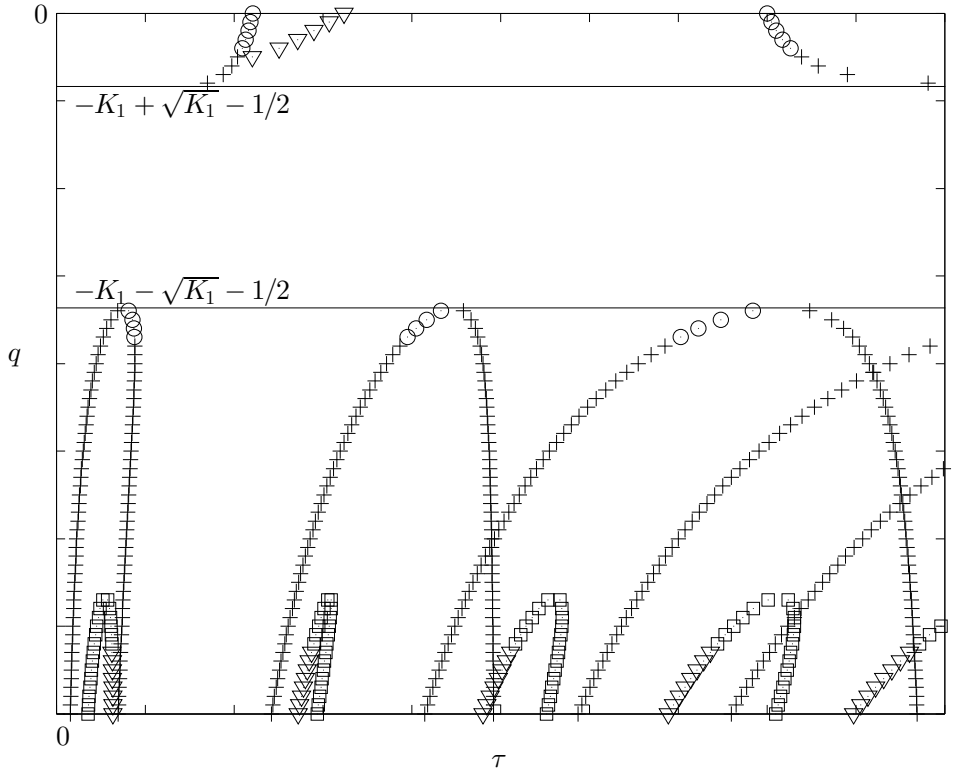}
  \caption{Hopf bifurcations for the curve \(L1:~r+q=-K_1\), with \(K_1>1\) in the plane \(\tau\times q\). For synchronous solutions, markers \('+'\) and \('\odot'\) denote bifurcations with \(\delta_s>0\) and \(\delta_s<0\), respectively; and for asynchronous solutions, markers \('\Box'\) and \('\bigtriangledown'\) denote \(\delta_a>0\) and \(\delta_a<0\).}
  \label{fig:PLL-application-example}
\end{figure}

For \(L1\) the boundaries for  \(q\) and \(r\) within this region are:
\begin{itemize}
\item with \(K_1\geq1\)
  \begin{align}
    \label{eq:boundaries-K1}
    \begin{array}{l}
q\in\left]-\left(K_1+\sqrt{K_1}+\dfrac{1}{2}\right),-\left(K_1-\sqrt{K_1}+\dfrac{1}{2}\right)\right[\\
      r\in\left]\dfrac{1}{2}-\sqrt{K_1},\dfrac{1}{2}+\sqrt{K_1}\right[
      \end{array}
  \end{align}
\item with \(K_1<1\)
  \begin{align}
    \label{eq:boundaries-K1-1}
    \begin{array}{l}      q\in\left]-\left(K_1+\sqrt{K_1}+\dfrac{1}{2}\right),-\dfrac{K_1}{2}\right[\\
      r\in\left]-\dfrac{K_1}{2},\dfrac{1}{2}+\sqrt{K_1}\right[.
      \end{array}
  \end{align}
  \end{itemize}
  For \(L2:\)
\begin{itemize}
\item with \(K_2\geq1\)
  \begin{align}
    \label{eq:boundaries-K2}
    \begin{array}{l}
q\in\left]-\left(K_2+\sqrt{K_2}+\dfrac{1}{2}\right),-\left(K_2-\sqrt{K_2}+\dfrac{1}{2}\right)\right[\\
      r\in\left]-\dfrac{1}{2}-\sqrt{K_2},-\dfrac{1}{2}+\sqrt{K_2}\right[
      \end{array}
  \end{align}
\item with \(K_2<1\)
  \begin{align}
    \label{eq:boundaries-K2-1}
    \begin{array}{l}      q\in\left]-\left(K_2+\sqrt{K_2}+\dfrac{1}{2}\right),-\dfrac{K_2}{2}\right[\\
      r\in\left]-\dfrac{K_2}{2},-\dfrac{1}{2}+\sqrt{K_2}\right[.
      \end{array}
  \end{align}
\end{itemize}
In Figure \ref{fig:PLL-application-example} are show curves of Hopf bifurcations in the plane \(\tau\times q\)  for the curve \(L1:~r+q=-K_1\), with \(K_1>1\), the time-delay and the direction of the bifurcations where computed  for synchronous solutions  using Eqs. \eqref{eq:freq-Pfix}, \eqref{eq:tau-pfix}, and \eqref{eq:delta-s}, and for asynchronous solutions, using \eqref{eq:freq-Pu} and \eqref{eq:delta-2}, the time-delay independent stability region is bounded by the horizontal lines at \(q=-(K_1\pm\sqrt{K_1}-1/2)\), as mention before, this stability criterion is conservative since, as it can be seen from the bifurcation curves, there exists regions outside  the limits bounded by Eq. \eqref{eq:boundaries-K1} where solutions still stable, although in this case its stability depends on the time-delay.

\section{Conclusions}
\label{sec:conclusions}
The condition \(f'_\phi(\phi^*)<0\), valid for first-order oscillator networks, is necessary but not sufficient to guaranty stability of both synchronous and asynchronous solutions for second-order networks. However, it is still possible to find a region with \(f'(\phi^*)<0\) bounded by functions of the normalized connection strength \(K\) (assuming \(b_0\) and \(\Omega\) are constants), where these solutions are time-delay independent stable. This analyses rely upon the assumption  \(f'_{\phi}(\phi^{*})\neq f'_{\phi_{\tau}}(\phi^{*})\) which allows the existence of non-degenerate Hopf bifurcations for synchronous solutions. This results are applicable to a vast variety of network configurations, from rings of two-ways oscillators to all-to-all networks, provided the \(S_N\)-symmetry is preserved.

\bibliographystyle{plain}

\end{document}